\documentclass[12pt,preprint]{aastex}

\slugcomment{Accepted for publication in the Astrophysical Journal}
\shortauthors{Hoard et al.}
\shorttitle{V592 Cas in the Mid-Infrared}

\begin{document}

\title{Observations of V592 Cassiopeiae with the {\em Spitzer
Space Telescope} -- Dust in the Mid-Infrared}
 
\author{D. W. Hoard\altaffilmark{1},
        Stella Kafka\altaffilmark{1},
        Stefanie Wachter\altaffilmark{1},
        Steve B. Howell\altaffilmark{2},
        Carolyn S. Brinkworth\altaffilmark{1},
        David R. Ciardi\altaffilmark{3},
        Paula Szkody\altaffilmark{4},
        Kunegunda Belle\altaffilmark{5},
        Cynthia Froning\altaffilmark{6},
        Gerard van Belle\altaffilmark{7}
}

\altaffiltext{1}{{\em Spitzer} Science Center, California Institute 
of Technology, Pasadena, CA 91125}
\altaffiltext{2}{National Optical Astronomy Observatory, Tucson, AZ 85719}
\altaffiltext{3}{NASA Exoplanet Science Institute, California Institute 
of Technology, Pasadena, CA 91125}
\altaffiltext{4}{Department of Astronomy, University of Washington, 
Box 351580, Seattle, WA 98195-1580}
\altaffiltext{5}{X-4 MS T085, Los Alamos National Laboratory, 
Los Alamos, NM 87545}
\altaffiltext{6}{Center for Astrophysics and Space Astronomy, 
University of Colorado, 593 UCB, Boulder, CO 80309}
\altaffiltext{7}{European Southern Observatory, 
Karl-Schwarzschild-Str.\ 2, 85748 Garching, Germany}

\begin{abstract}
We present the ultraviolet--optical--infrared spectral energy 
distribution of the 
low inclination novalike cataclysmic variable V592 Cassiopeiae, 
including new mid-infrared observations from 3.5--24 $\mu$m 
obtained with the {\em Spitzer Space Telescope}.
At wavelengths shortward of 8 $\mu$m, the spectral energy 
distribution of V592 Cas is dominated by the steady state 
accretion disk, but there is flux density in excess of the 
summed stellar components and accretion disk at longer wavelengths. 
Reproducing the observed spectral energy distribution from 
ultraviolet to mid-infrared wavelengths can be accomplished by 
including a circumbinary disk composed of cool dust, 
with a maximum inner edge temperature of $\approx500$ K.
The total mass of circumbinary dust in V592 Cas ($\sim10^{21}$ g) 
is similar to that found from recent studies of infrared excess 
in magnetic CVs, and 
is too small to have a significant effect on the 
long-term secular evolution of the cataclysmic variable.  
The existence of circumbinary dust in V592 Cas is possibly linked 
to the presence of a wind outflow in this system, 
which can provide the necessary 
raw materials to replenish the circumbinary disk on 
relatively short timescales, and/or could be a remnant from the 
common envelope phase early in the formation history of the system.
\end{abstract}

\keywords{accretion, accretion disks --- 
novae, cataclysmic variables --- stars: individual (V592 Cas)}

\section{Introduction}

V592 Cassiopeiae was discovered by \citet{greenstein70}.  
Its blue color and spectral characteristics identified it as a 
disk-accreting novalike cataclysmic variable (CV). 
Its optical spectrum displays strong \ion{He}{2} 4686 \AA\ and 
\ion{C}{3}/\ion{N}{3} Bowen blend emission, with weak Balmer 
and \ion{He}{1} emission lines \citep{downes95}. 
\citet{huber98} performed the first time-resolved optical 
spectroscopic study of V592 Cas and used radial velocity 
measurements to determine an
initial estimate of the orbital period, $P_{\rm orb}=2.47$ hr. 
This was later refined to $P_{\rm orb}=2.76$ hr by \citet{taylor98} 
and \citet{witherick03}.
Although photometric observations showed that it does not eclipse 
(orbital inclination of $i=28^{\circ}$; \citealt{huber98}), 
V592 Cas does show both positive and negative 
permanent superhumps \citep{taylor98}.
Optical and ultraviolet (UV) spectroscopic observations demonstrated 
the presence of a wind, which manifests primarily as a P Cygni 
profile in the emission lines \citep{witherick03,prinja04,kafka08}.
\citet{kafka08} showed that the wind is variable and 
reaches velocities up to $v_{\rm wind}\sim5000$ km s$^{-1}$.  

The secular evolution of interacting binary stars is driven
by the transfer of matter between the stellar components and 
out of the system, and the corresponding redistribution and 
loss of angular momentum.
However, the observed orbital period distribution of the CV 
population (which provides a snapshot of the evolutionary 
process) shows several discrepancies compared to  
the standard two-mechanism (gravitational radiation and 
magnetic braking from the secondary star) model for angular momentum loss.  
For example, population synthesis studies typically predict 
an orbital period minimum that is significantly shorter 
($\sim65$--$70$ min; e.g., \citealt{patterson98,howell01,willems05}) 
than the observed minimum of $\sim75$--$85$ min 
\citep{patterson98,knigge06}.  Also, the standard two mechanism 
model predicts a tight relationship between orbital period and 
mass transfer rate, whereas the CV population shows a spread of 
up to several orders of magnitude in mass transfer rate at a 
given orbital period \citep{patterson84,kolb01}.  
An additional angular momentum 
loss mechanism, which would cause CVs to reach a minimum 
orbital separation (and, hence, period) sooner during their 
secular evolution and contribute to the spread in mass transfer 
rates, could account for this.

In this paper, we use new infrared (IR) observations of V592 Cas 
obtained from the {\em Spitzer Space Telescope} to explore 
a mechanism that has been proposed as an additional source of 
angular momentum loss in CVs:\ gravitational torques on the 
inner binary from a circumbinary disk, likely composed of cool dust 
\citep{spruit01,taam03,willems05,willems07}.
In population simulations, the presence of a circumbinary disk 
can be at least as 
important as magnetic braking in the secular evolution of a 
CV, and can lead to completely dissolving the secondary star 
in a finite time, at 
an age $\sim3$ times smaller than in an evolutionary model 
that does not include a circumbinary disk \citep{spruit01}.
In our previous work, we have used {\em Spitzer} mid-IR 
observations of 
several short orbital period ($P_{\rm orb}=1.3$--1.8 hr) 
magnetic CVs \citep{howell06,brinkworth07,hoard07}
to demonstrate an infrared excess
caused by the presence of cool ($T\lesssim1000$ K), 
likely circumbinary, dust.  In all cases, 
the total mass of dust required to reproduce the 
observed mid-IR spectral energy distributions (SEDs) was 
much smaller than the amount predicted to be necessary to 
have a significant effect on CV evolution.  We now present 
the first analysis of {\em Spitzer} observations of a bright 
novalike (non-magnetic) CV at a longer orbital period, 
which offers a more 
likely environment for a substantial dust presence.

\section{Observations and Data Processing}

We observed V592 Cas with each of the three instruments 
on-board {\em Spitzer} \citep{werner04} during the GO2 cycle.  
Details of the observations and data processing are given below.

\subsection{Infrared Array Camera}
\label{s:obs-IRAC}

We obtained photometric measurements in each of the four imaging 
channels\footnote{Channels 1--4 are referred to by corresponding 
wavelengths of 3.6, 4.5, 5.8, and 8.0 $\mu$m; however, 
see \S\ref{s:irac_errs} and
Table \ref{t:data}.} of the Infrared Array Camera 
(IRAC; \citealt{fazio04}) using 5 medium scale, cycling 
dithers of the 12-second frame time, which are 
available as standard options 
of the IRAC imaging Astronomical Observing Request (AOR) template.  
The observation (AOR identification number 
14408960) was executed on 18 Aug 2005 UT as part of IRAC 
campaign 24.  The data were processed through the 
{\em Spitzer} Science Center (SSC) S14.0.0 
pipeline to yield basic calibrated data (BCD) images.  
We applied the IRAC array location dependence correction 
to these BCDs, as described in the IRAC Data Handbook\footnote{See 
\url{http://ssc.spitzer.caltech.edu/irac/dh/}.}.

We did not perform photometry on a mosaicked image created 
from the BCDs because of the poor results we found when 
attempting this in previous work \citep{brinkworth07}.
Instead, we performed aperture photometry on the individual 
corrected BCD images using the IRAF\footnote{The Image 
Reduction and Analysis 
Facility is maintained and distributed by the National Optical 
Astronomy Observatory.} task {\sc qphot}.  We utilized a 3-pixel 
radius aperture (1 pixel $\approx$ 1.22\arcsec), with a 
3--7 pixel background annulus, which is one of the 
configurations for which aperture corrections are available 
in the IRAC Data Handbook (factors of 
1.124, 1.127, 1.143, and 1.234 for channels 1--4, respectively).  

The individual IRAC channel 1 BCD images display a column artifact 
caused by a bright, saturated field star near the edge of the 
field of view.  This column passes through the edge of the 
V592 Cas stellar profile, near the first diffraction ring.
We investigated correcting for this artifact by interpolating 
across adjacent pixels.
However, it became clear that the likely effect of this 
artifact is to decrease the nominal value of the channel 1 
photometric measurement by less than $1\sigma$, so we did not 
include the correction to the final photometric measurement
used in our SED of V592 Cas.
The BCDs from the other IRAC channels do not contain this 
artifact because the bright star is not saturated.

As described in the IRAC Data Handbook, we applied the pixel 
phase dependence correction to the channel 1 (3.6 $\mu$m) 
photometry.  However, we did not perform the color correction other 
than to utilize the isophotal effective channel wavelengths 
(see Table \ref{t:data}) during our subsequent interpretation 
of the data, which accounts for all but a few per cent of the 
color correction.  The remaining effect of the color correction is 
folded into our IRAC systematic uncertainty budget -- see below.  
We then performed a $5\sigma$ clipping rejection\footnote{That 
is, a given flux density is compared to the average and standard 
deviation ($\sigma)$ of all of the {\em other} flux density 
values.  The measurement is rejected if it is more than $5\sigma$ 
away from the average.} on the set of 5 individual BCD flux 
densities, which resulted in the rejection of only 1 measurement, 
from the channel 3 (5.8 $\mu$m) data.  Finally, the individual 
BCD photometry values for each channel were averaged together.

\subsubsection{Uncertainties in the IRAC Photometry}
\label{s:irac_errs}

\begin{figure}
\epsscale{0.95}
\plotone{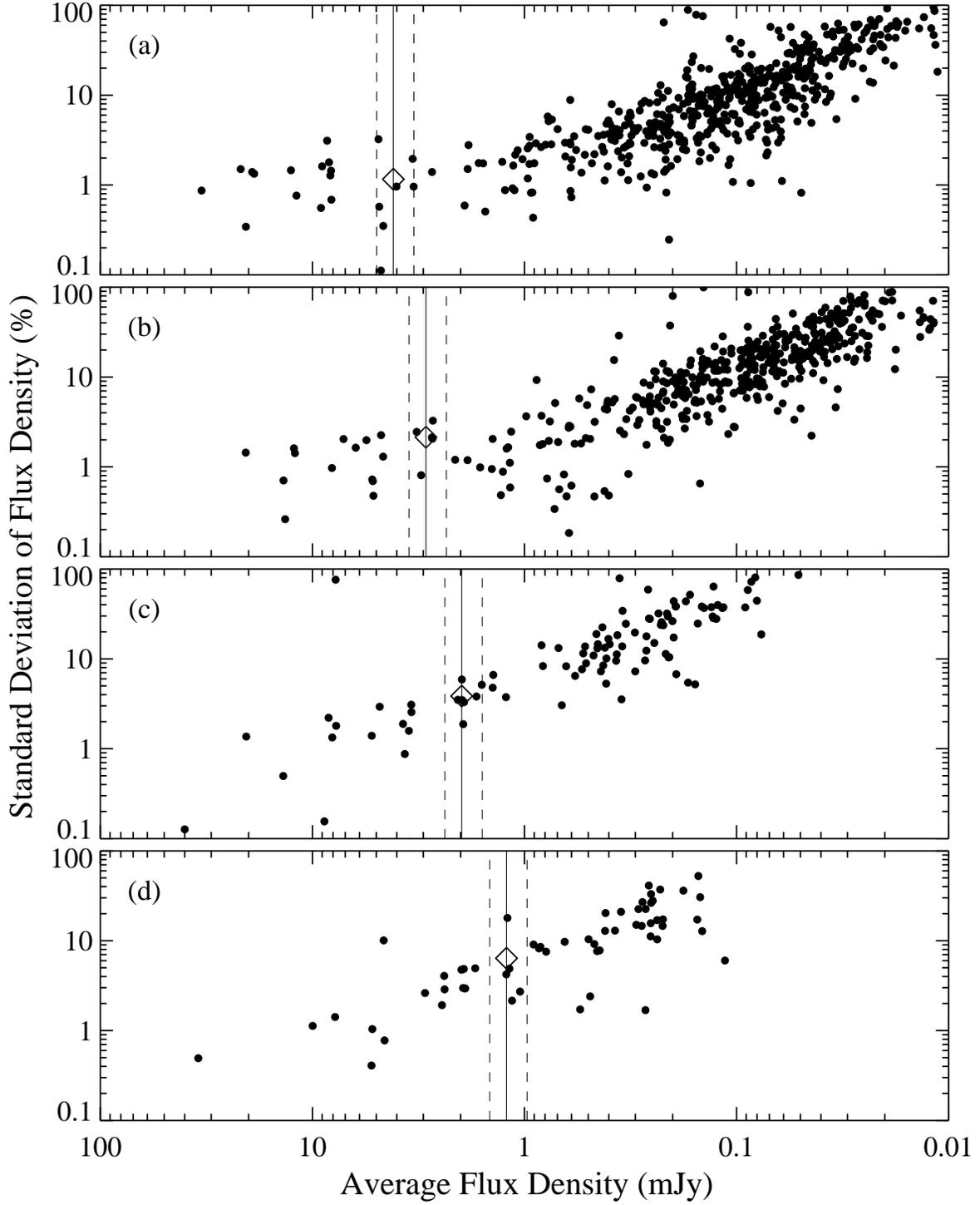}
\epsscale{1.00}
\caption{Per cent uncertainty (standard deviation) of 
measured IRAC flux densities as a function of average 
source flux density.  The panels show (a) channel 1 
(3.6 $\mu$m), (b) channel 2 (4.5 $\mu$m), (c) channel 3 
(5.8 $\mu$m), and (d) channel 4 (8.0 $\mu$m).  
The solid vertical line in each panel is the mean flux 
density of V592 Cas in the corresponding channel.  
The vertical dashed lines show the $\pm20$\% range over 
which uncertainty measurements were averaged to arrive at the 
scatter term uncertainties for V592 Cas (large diamonds).  
See \S\ref{s:irac_errs} for details.
\label{f:errs}}
\end{figure}

For IRAC photometry, the total uncertainty budget includes 
several systematic terms:\ 3\% for the absolute gain calibration, 
1\% for repeatability, 3\% for the absolute flux 
calibration of the IRAC calibration stars, and 1\% for the 
color correction.  These values are 
from version 3.0 of the IRAC Data Handbook.  In addition, 
there is a scatter term, which we have determined as follows 
(based on the method described in \citealt{brinkworth07}).  
We performed additional aperture photometry on the BCDs, as described 
above, for all point sources (except V592 Cas) in the mutually 
overlapping regions of each IRAC channel in each of the 5 
dithered exposures from both the on- and off-target fields of 
the two IRAC channel pairs (i.e., resulting in a total of 10 
sets of photometry for each channel, comprised of up to 5 
measurements of each object in two different fields).  
We applied the $5\sigma$ clipping rejection to the photometry, 
and kept all objects for which 3 or more (out of a maximum 
of 5 possible) measurements were available.  
This left a total of 543, 446, 112, and 62 objects in 
channels 1--4, respectively.  We then plotted, for each 
channel, the standard deviation of the individual flux 
densities for each object as a function of its average flux 
density.  This results in well-behaved correlations between 
the scatter and flux density in each channel, which we show 
in Figure \ref{f:errs} for future use.  We then determined 
what we call the ``field-scatter'' uncertainty for V592 Cas 
by averaging the points on these plots within $\pm20$\% of 
the mean flux density in each channel.  This results in 
field-scatter terms of 0.05 (1.2\%), 0.06 (2.1\%), 0.08 (3.9\%), 
and 0.08 (6.4\%) mJy in channels 1--4, respectively.  
The field-scatter term is then added to the total systematic 
uncertainty budget described above.  This process is more 
robust than simply using the ``self-scatter'' term 
(i.e., the standard deviation of the individual BCD photometry 
values for V592 Cas itself) since it is based on the results 
from a large number of objects and is not influenced by any 
intrinsic variability of the CV.

\subsection{Infrared Spectrograph Peak-Up Imaging Array}
\label{s:obs-IRS}

We obtained a single photometric measurement with the blue Peak-Up 
Imaging (PUI) array of the Infrared Spectrograph (IRS; 
\citealt{houck04}) using 30 small scale, cycling dithers of 
the 30-second ramp duration, 
which are available as standard options in the 
IRS PUI science mode AOR template.  
The observation (AOR identification number 14412032) 
was executed on 16 Jan 2006 UT 
as part of IRS campaign 28.  The data were processed through 
the SSC S15.3.0 pipeline to yield basic calibrated data (BCD) 
images and a single post-BCD mosaic image.  

We performed aperture photometry on the post-BCD mosaic using 
the {\sc qphot} task in IRAF.  In order to avoid contamination from 
the relatively bright nearby star 2MASS J00205373+5542192 
(star 14 in \citealt{henden95}, located 13\arcsec\ east), 
which is the only other object detected in the IRS-PUI mosaic, 
we utilized a 3-pixel radius aperture (1 pixel = 1.8\arcsec), 
with the background computed from a mean over the entire image 
(excluding the two stars visible in the frame).  
The aperture correction for this configuration (factor of 1.56) 
is available in the IRS Data Handbook\footnote{See 
\url{http://ssc.spitzer.caltech.edu/irs/dh/}.}.  
As with the IRAC data, 
we did not perform the color correction, other than to utilize 
the isophotal channel wavelength (see Table \ref{t:data}) 
during our subsequent interpretation of the data.  The 
remaining effect 
of the color correction is folded into our IRS-PUI 
uncertainty budget (see below).

\subsubsection{Uncertainties in the IRS-PUI Photometry}
\label{s:irs_errs}

For IRS-PUI photometry, the total systematic uncertainty budget 
is the quadrature sum of 2\% for the flux density zero point, 3\% for 
repeatability, 3.4\% for deviation from blackbody source SED 
in the aperture correction, and 5\% for (reasonable) deviation 
from a flat SED in the color correction.  These values reflect 
upper limits from version 3.2 of the IRS Data Handbook.  
In addition, the total uncertainty includes a scatter term 
of 0.07 mJy (12.5\%) obtained for the target aperture in 
the uncertainty (``unc'') image provided by the standard
processing pipeline for the post-BCD mosaic.  

Because only two sources are visible in the small field-of-view 
of the IRS-PUI image, we could not repeat the field-scatter 
determination used with the IRAC data.  We note that the average 
flux density of V592 Cas obtained by performing photometry on 
each of the 30 individual BCD images is $0.56\pm0.12$ mJy 
(where the uncertainty is the self-scatter term only), which 
is in agreement with the value measured from the post-BCD 
mosaic (see Table \ref{t:data}).  The larger scatter term in 
this case (which amounts to 21\% of the measured flux density) 
is likely a result of the fact that V592 Cas is very faint in 
any individual BCD, resulting in low S/N individual measurements.  
By comparison, the other source in the IRS-PUI field is brighter, 
with a mean flux density of 1.92 mJy and a scatter uncertainty 
amounting to only 7.5\%.

\subsection{Multiband Imaging Photometer for {\em Spitzer}}
\label{s:obs-MIPS}

We obtained a single photometric measurement with the 24 $\mu$m 
array of the Multiband Imaging Photometer for {\em Spitzer} 
(MIPS; \citealt{rieke04}) using 20 cycles of the 10-second 
exposure time small field photometry sequence, which are
available as standard options in the MIPS pointed imaging 
AOR template.  The observation
(AOR identification number 14412288) 
was executed on 21 Feb 2006 UT as part 
of MIPS campaign 29.  The data were processed through the 
SSC S16.0.1 pipeline to yield basic calibrated data (BCD) images 
and a single post-BCD mosaic image.  We re-mosaicked the data 
using MOPEX (version 16)\footnote{See 
\url{http://ssc.spitzer.caltech.edu/postbcd/onlinedocs/index.html}.} 
with overlap correction turned on, after first performing a 
self-calibration of the images; that is, we created a median 
image from all of the BCDs, normalized it to a value of 1.0, 
then divided all of the BCDs by that median image.  This procedure 
removes persistent image artifacts that are not removed during the 
standard pipeline processing, such as long-term latents, 
low-level ``jailbars'' (that is, parallel linear artifacts produced 
following the read-out of a saturated source in the field), 
and gradients (see the MIPS Data Handbook\footnote{See 
\url{http://ssc.spitzer.caltech.edu/mips/dh/}} for details of the 
self-calibration procedure and MIPS image artifacts).
We resampled the pixel 
scale in this mosaic to 1.225 arcsec pixel$^{-1}$ (i.e., 
oversampled by a factor of 2 compared to the standard post-BCD 
mosaic).

We performed aperture photometry on our mosaic using the {\sc qphot} 
task in IRAF.  We utilized a 7-arcsecond radius aperture, with 
a 20--32-arcsecond background annulus.  The aperture correction 
for this configuration (factor of 1.61) is available in the 
MIPS Data Handbook.  
We did not utilize the standard 7--13-arcsecond background 
annulus in order to exclude the star located 13\arcsec\ east 
of V592 Cas (see \S\ref{s:obs-IRS}). 
As with the data from the other instruments, we did not perform 
the color correction other than to utilize 
the isophotal channel wavelength (see Table \ref{t:data}) 
during our subsequent interpretation of the data; the remaining effect 
of the color correction is folded into our MIPS 
uncertainty budget (see below).

\subsubsection{Uncertainties in the MIPS Photometry}
\label{s:mips_errs}

For MIPS photometry, the total systematic uncertainty budget 
is the quadrature sum of 4\% for the absolute calibration, 
0.4\% for repeatability, 
and 3\% for the color correction.  These values reflect upper 
limits from version 3.3 of the MIPS Data Handbook.  In addition, 
the total uncertainty includes a scatter term of 0.06 mJy (14\%) 
obtained for the target aperture in 
the uncertainty (``munc'') image provided 
by MOPEX for our mosaic.  

Because only a few bright sources are visible in the individual 
MIPS BCD images, we could not repeat the field-scatter 
determination used with the IRAC data, nor could we perform 
photometry for V592 Cas itself in the individual BCDs.  
We note that the average flux density of V592 Cas obtained 
by performing photometry on the standard post-BCD pipeline 
mosaic image is $0.45\pm0.06$ mJy (where the uncertainty is 
the scatter term only), which is consistent with the result 
from our improved mosaic (see Table \ref{t:data}).  
The detailed characterization of the properties of 
MIPS 24 $\mu$m photometry presented in \citet{vonbraun08} 
suggests a somewhat lower fractional scatter uncertainty 
should be obtained for a source with the flux density of 
V592 Cas, in the range 7--13\% (see Figure 6 in that work).  
However, the slightly larger fractional scatter uncertainty 
that we observe in our MIPS photometry (14\%) likely results 
from the fact that our single photometric measurement was obtained 
from a mosaic created from 280 individual BCD images 
(whereas each photometric measurement shown in Figure 6 of 
\citealt{vonbraun08} was obtained from the average 
of 1224--2448 individual BCDs) coupled with the slightly 
higher 24 $\mu$m background during the MIPS observations 
of V592 Cas (22.9 MJy sr$^{-1}$) compared to GU Boo 
(19.7--21.0 MJy sr$^{-1}$)\footnote{We note that one-half 
of the GU Boo MIPS data presented in \citet{vonbraun08} 
were obtained during the same MIPS instrument campaign, 
and on the same day, as our MIPS observation of V592 Cas.}.  
With that in mind, our MIPS photometry results are consistent 
with the expected level of photometric uncertainty found 
by \citet{vonbraun08} for a source of this mean brightness.

\section{Spectral Energy Distribution Model}
\subsection{The Data}
\label{s:data}

Table \ref{t:data} lists our observed {\em Spitzer} data for V592 Cas, 
as well as adopted flux densities at shorter wavelengths.
The latter consists of 
two UV photometric measurements from an {\em International 
Ultraviolet Explorer} ({\em IUE}) spectrum (as shown in Figure 1 of 
\citealt{witherick03}; also see \citealt{guinan82}), and ground-based 
$V$ (see below) and 2MASS near-IR $J$, 
$H$, and $K_{\rm s}$ photometric measurements.

When constructing an SED from these data, we do not have to be 
concerned about orbital variations since V592 Cas is a low 
inclination, non-eclipsing system.  However, we do have to assume 
that non-coherent variability on the order of a couple tenths of 
a magnitude (i.e., normal CV ``flickering'') is present, as well 
as consider the possibility that there could be long-term trends 
in the brightness of V592 Cas.
To aid in this assessment, in Table \ref{t:oldphot} we present a 
compilation of published optical and near-IR photometry of 
V592 Cas that is, to the best of our knowledge, 
complete\footnote{Some photometry results, particularly the 
$UBV$ measurements from \citet{haug70}, are repeated in multiple 
references.  In such cases, we have only presented the 
original reference source here.}.

The AAVSO measurement was obtained contemporaneously with 
our {\em Spitzer} observations, and shows that V592 Cas was 
relatively constant in brightness during that time.  
The most recent $V$-band measurement ($\langle V \rangle=12.55$; 
\citealt{kafka08}), which was obtained $\approx2$ months prior to 
the IRAC observation, is somewhat brighter than historic 
observations of V592 Cas
($V\approx12.7$--$12.8$; \citealt{haug70,huber98,taylor98}). 
However, \citet{zwitter94} determined a color index of $(V-R)=+0.11$ 
for V592 Cas from their spectroscopic observations, so the AAVSO 
measurement corresponds to $V\approx12.4$, which is consistent 
within the range of uncertainty and flickering variability of the 
$V$-band measurement from \citet{kafka08}.  
Consequently, we have adopted the \citet{kafka08} value 
of $V=12.55$ for our SED.
The only available near-IR photometry other than that from 2MASS 
appears to have suffered from calibration problems (see note 
in Table \ref{t:oldphot}), so we have used  
the 2MASS photometry values 
(in any case, these measurements do not appear to be 
significantly discrepant in the SED compared to the data at 
other wavelengths).
The optical photometry was converted from magnitudes to 
flux densities using the zero points given in \citet{cox00}; 
for the
2MASS All Sky Point Source Catalog photometry we used the 
zero points in \citet{cohen03}.  

The flux density uncertainties for the 2MASS data are the $1\sigma$ 
photometric uncertainties propagated through the corresponding 
magnitude to flux density conversions.  For the ground-based 
optical data, the uncertainty reflects the observed range of $V$ 
magnitude from \citet{kafka08} propagated through 
the flux density conversion.  
The flux density uncertainty on the UV measurements, which were read 
from a plot of the spectrum, is estimated at 10\%.
Derivation of the flux density uncertainties for the {\em Spitzer} 
data from each instrument is described in \S\ref{s:irac_errs}, 
\S\ref{s:irs_errs}, and \S\ref{s:mips_errs}.  In all cases, 
the ``uncertainties'' on the filter wavelengths represent the 
widths of the photometric bands.

\subsubsection{Reddening and Distance to V592 Cas}
\label{s:dist}

\citet{nandy75} -- among others -- derive a correlation 
of the equivalent width of the strong UV interstellar 
absorption band at 2200 \AA\ with optical color excess, $E(B-V)$, 
and the band profile shape with the nature and size of 
the grains producing the observed extinction.  
In this manner, \citet{guinan82} used the 
{\em IUE} spectrum of V592 Cas from 5 Dec 1981 UT to
obtain a color excess (reddening) of $E(B-V)=0.3$.  
\citet{taylor98} used observations of the stationary 
(i.e., interstellar) Na D absorption in V592 Cas to 
derive a lower reddening of $E(B-V)=0.07$--$0.20$.
They adopt a ``compromise'' value with the \citet{guinan82} 
result of $E(B-V)=0.25$.

In the optical spectra of V592 Cas presented in \citet{kafka08}, 
the Na D lines are contaminated by Tucson city light emission, 
so we could not repeat the \citet{taylor98} measurement.
However, we reanalyzed the archival {\em IUE} spectrum of V592 Cas, by
applying the IRAF task {\sc deredden}, which utilizes the 
empirical extinction law of \citet{cardelli89}.
Removal of the 2200 \AA\ absorption feature required a 
reddening of $E(B-V)\approx0.22$.  This is consistent 
with the \citet{taylor98} value.  

\begin{figure}
\epsscale{0.95}
\plotone{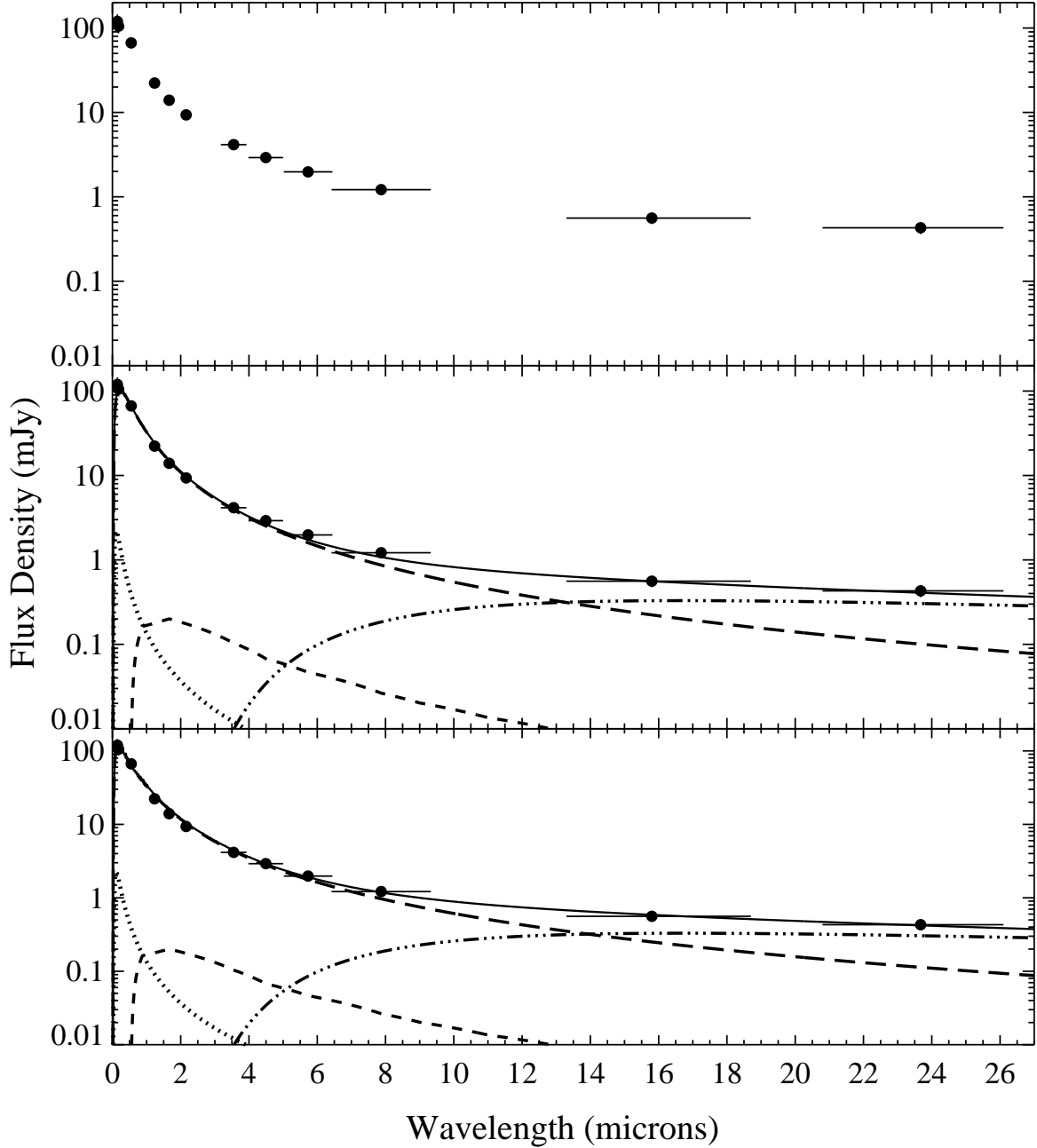}
\epsscale{1.00}
\caption{Dereddened spectral energy distribution (top panel) 
and model (middle and bottom panels) for V592 Cas.
The photometric data (see \S\ref{s:data}) are shown as filled circles.
The system model (solid line) is composed 
of a WD (dotted line), M5.0 secondary star (short dashed line), 
steady state accretion disk (long dashed line), and 
circumbinary dust disk (dot-dot-dot dash line).  
The model in the middle panel utilizes a limb darkened flat 
accretion disk with no irradiation, while the model in the 
bottom panel utilizes a limb-darkened flared accretion disk 
with irradiation.
See \S\ref{s:models} for a discussion of the model parameters.
\label{f:sed}}
\end{figure}

We used our reddening value with the diffuse interstellar 
extinction law from \citet{fitzpatrick07} to deredden the 
observed flux densities of V592 Cas.  
Because there is some uncertainty associated with the extinction 
curve (e.g., see Figure 9 in \citealt{fitzpatrick07}), the 
dereddening process increases the uncertainties of the dereddened 
flux densities relative to the corresponding observed flux densities.  
The reddening correction is largest at short wavelengths and 
becomes increasingly negligible in the IR.  Consequently, to avoid 
unnecessarily degrading the accuracy of our observed flux 
densities, we did not apply the dereddening correction to any 
photometric measurement for which the fractional change produced by 
the correction would be smaller than 25\% of the fractional 
standard deviation of the observed measurement 
($\sigma_{\rm f_{\nu}}/f_{\nu}$).  In practice, this means that 
the dereddening correction was not applied to any of the 
{\em Spitzer} data -- see Table \ref{t:data}.
Figure \ref{f:sed} (top) shows the dereddened SED for V592 Cas 
and the corresponding dereddened flux densities are 
listed in Table \ref{t:data}.

\citet{harrison04} have updated the $M_{\rm V}$--$P_{\rm orb}$ 
relation for dwarf novae in outburst \citep{warner87} by using 
distances from {\em Hubble Space Telescope} Fine Guidance Sensor 
trigonometric parallax measurements.  They have also shown that 
this relation appears to be applicable to novalike CVs as well.  
We used this relation, with extinction appropriate for our 
reddening value ($A_{\rm V}=0.66$ mag), to determine an expected 
(dereddened) absolute magnitude for V592 Cas of $M_{\rm V,corr}=4.9$ 
or $M_{\rm V,obs}=4.1$, where the former is the inclination-normalized 
value (see description in \citealt{harrison04} and references 
therein) and the latter is the value that would be observed at 
the nominal inclination of V592 Cas ($i=28^{\circ}$).  
Using $V=12.55$, this gives a distance to V592 Cas of 360 pc.  
\citet{taylor98} made a similar calculation using the old 
calibration of the $M_{\rm V}$--$P_{\rm orb}$ relation, and 
found distances of 350 pc (if interstellar extinction is 
ignored), 240 pc (including extinction), or their final adopted 
distance of 330 pc (assuming that V592 Cas is ``a bit more 
luminous than dwarf novae'').  It is not clear if they made the 
correction for inclination, which would have been of the same 
amplitude and direction as that produced by their assumption 
that V592 Cas was more luminous than an outbursting dwarf nova.  
In addition, the revision to the calibration of the 
$M_{\rm V}$--$P_{\rm orb}$ relation would also have acted to 
make their distance larger by about 3\%.  We have scaled all 
of our SED model components (see \S\ref{s:models}) to a 
distance of 360 pc.

\subsection{The Model}
\label{s:models}

\begin{figure}
\epsscale{1.00}
\plotone{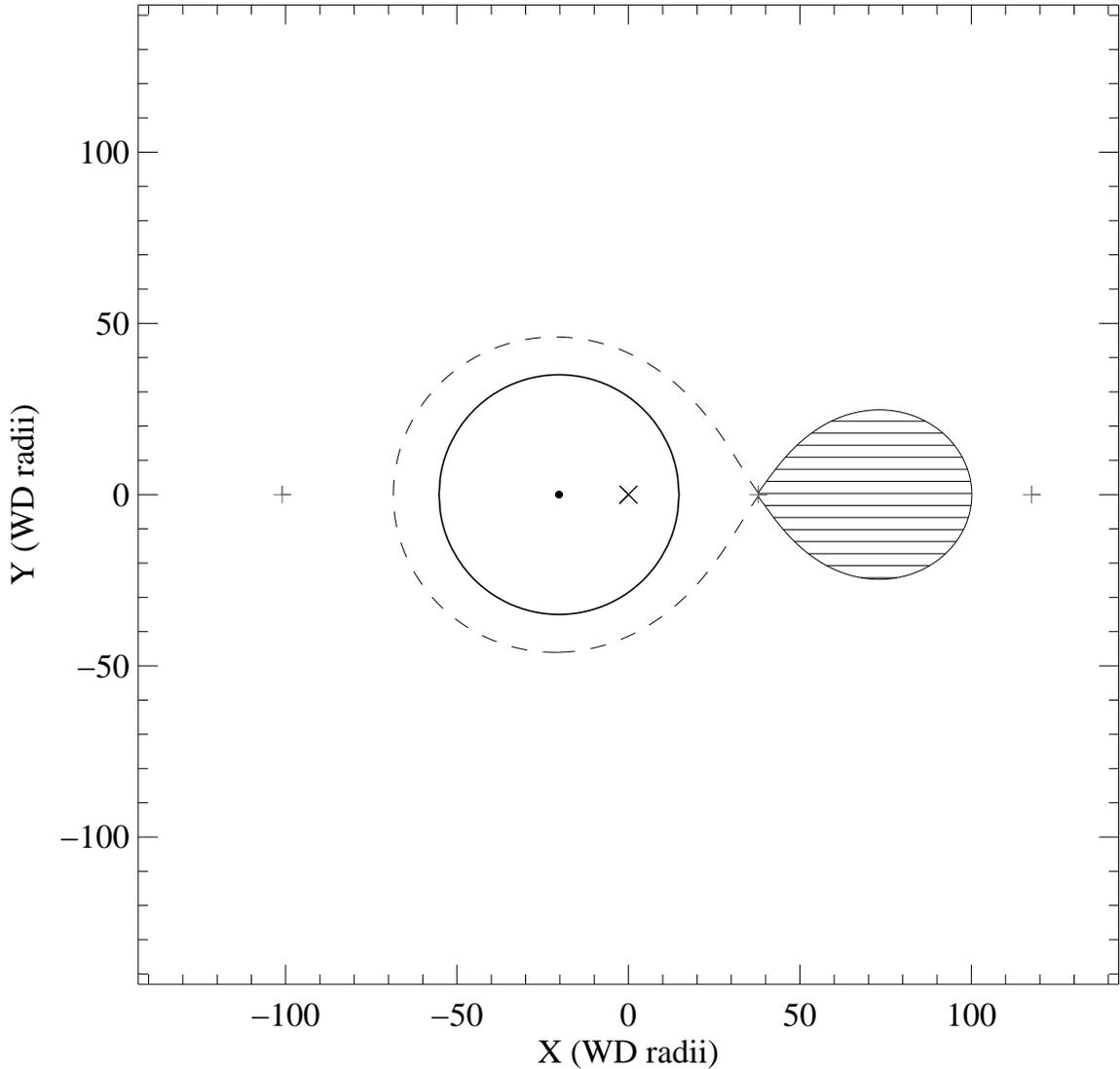}
\epsscale{1.00}
\caption{To-scale diagram of V592 Cas, including the model 
components (based on the parameters listed in Table \ref{t:model} 
and discussed in the text).  The diagram shows the secondary 
star (horizontal hatched area), WD (small filled circle), 
WD Roche lobe (dashed line), and the accretion disk (solid 
line around the WD marks the outer edge).  The circumbinary 
dust disk, which has an inner radius of 700 $R_{\rm wd}$, 
is not shown.  The inner and outer Lagrange points 
(plus symbols) and system center of mass (cross symbol) 
are also shown.  The CV is depicted as viewed from ``above'' 
(i.e., with the orbital plane in the plane of the sky, 
equivalent to a system inclination of $0^{\circ}$).
\label{f:geom}}
\end{figure}

Our custom-built IR SED modeling code for CVs is
described in \citet{brinkworth07} and \citet{hoard07}.
We have continued to improve the code by implementing a model 
component to represent the accretion disk in non-magnetic CVs 
(described below).  
For this work, we have used a model consisting of a WD (denoted 
with subscript ``wd''), secondary star (subscript ``ss''), 
optically thick steady state accretion disk (subscript ``acd''), 
and optically thin circumbinary dust disk (subscript ``cbd'') 
to compare to the observed SED of V592 Cas.  The system and 
model parameters are listed in Table \ref{t:model}, and 
Figure \ref{f:sed} (bottom) shows the SED data for V592 Cas 
with the nominal model.   Figure \ref{f:geom} shows a to-scale 
diagram of V592 Cas based on the nominal model.  We discuss 
some details of the model components and parameter choices below.

\subsubsection{The White Dwarf}

The WD is represented by a blackbody curve with 
$T_{\rm wd} = 45,000$ K, 
which was chosen as a ``typical'' WD temperature for 
novalike CVs (e.g., \citealt{araujo03,hoard04}).  
We have kept the WD temperature fixed at this value throughout our
model calculations; for example, ignoring the possible role of the
rate of mass accretion in heating the WD.
In any case, as described below, even at this temperature the WD is only 
a minor contributor to the total system flux distribution 
longward of the ultraviolet (UV).  If the temperature of the 
WD in V592 Cas is significantly different than this, then it 
is most likely to be smaller \citep{sion99}, which would make 
the WD even less significant in the determination of a model 
that can reproduce the long wavelength observations.  

The WD is assumed to have the mean WD mass in CVs with $P_{\rm orb}<6$ hr, 
$M_{\rm wd}=0.75 \, M_{\odot}$ \citep{knigge06}, and a 
corresponding radius of $R_{\rm wd}=0.0106 \, R_{\odot}$ 
\citep{hamada61,panei00}.
The projected surface area of the WD has been reduced by 7\%, 
consistent with obscuration by an accretion disk (see below) 
that extends outward from $R_{\rm acd,in} = 1 R_{\rm wd}$ 
at the system inclination of $i=28^{\circ}$ \citep{huber98}.

\subsubsection{The Secondary Star}

The empirical CV secondary star sequences of \citet{smith98} 
and \citet{knigge06} predict a secondary star spectral type 
of M4.5$\pm$1 V (their Equation 4) and M4$\pm$2 V (his 
Figure 9), respectively, for the 2.76 hr orbital period 
of V592 Cas.  
That is, regardless of the true nature of the secondary stars 
in CVs, which likely differ in internal structure from 
corresponding single main sequence stars (e.g., by being 
oversized and/or slightly evolved as a result of their history 
of mass loss and irradiation -- see discussion in \citealt{knigge06} 
and references therein)
and/or by being non-uniformly heated by irradiation from the 
WD and accretion disk, the secondary star in V592 Cas has the 
spectroscopic appearance of a main sequence star with spectral type 
in the range indicated above.
As it turns out (see Figure \ref{f:sed}), the nominal secondary star 
contributes $\lesssim2.5$\% of the total flux density at all 
wavelengths, so the choice of secondary star template is robust 
against significant changes in spectral type or overall 
stellar temperature/luminosity that might be produced by evolutionary 
or irradiation effects (e.g., changing to an M2 template increases 
the maximum secondary star contribution to only $\sim4$\% of the 
total SED).

Thus, to represent the secondary star, we have used an empirical 
template of an M5.0 V star (GJ1093) constructed from optical 
\citep{mermilliod86,monet03}, 2MASS near-IR, and {\em Spitzer} 
IRAC \citep{patten06} photometry. The secondary star template 
was extended to wavelengths longer than 8 $\mu$m by 
extrapolating the shorter wavelength data along a 
Rayleigh-Jeans tail (i.e., $f_{\nu} \propto \lambda^{-2}$).  
The secondary star template was scaled from the trigonometric 
parallax distance of $d=7.76$ pc for GJ1093 \citep{oppenheimer01} 
to the adopted distance of V592 Cas, $d=360$ pc 
(see \S\ref{s:dist}).  Average mass, initial radius\footnote{The 
secondary star size and shape were recalculated within the Roche 
geometry of the binary in our model -- see Figure \ref{f:geom}.}, 
and temperature values for a star of this spectral type (compiled 
in \citealt{cox00}) were used in the model.

\subsubsection{The Accretion Disk}

The accretion disk component assumes an optically thick steady 
state disk composed of concentric rings emitting as blackbodies, 
following the ``standard model'' prescription in \citet{FKR}.  
The accretion disk component is parameterized by 
inner and outer radii 
($R_{\rm acd, in}$ and $R_{\rm acd, out}$), height at the inner
radius ($h_{\rm acd}$), and a mass transfer 
rate from the secondary star ($\dot{M}$).  
The inclination of the disk is assumed to be the same as the 
system inclination; higher inclination systems have smaller 
projected surface area, resulting in an overall decrease in 
the effective brightness of the disk. 
In the case of V592 Cas, the inclination is quite low 
($i=28^{\circ}$) so obscuration of the WD 
by the disk (and vice-versa), although included in our 
model calculation, has an insignificant effect on the 
resultant SED.

The general shape of the model accretion disk SED 
is similar to a slightly flattened and stretched blackbody 
function; that is, it displays a relatively broad peak with 
a rapid drop in brightness on the short wavelength side and 
a gradual (Rayleigh-Jeans-like) decline on the long wavelength 
side (e.g., see Figure 20 in \citealt{FKR}).  In general, 
increasing either the disk area (i.e., making $R_{\rm acd,out}$ 
larger) or the mass transfer rate causes the overall accretion 
disk SED to become brighter.  Increasing $R_{\rm acd,out}$ also 
tends to shift the peak of the accretion disk SED toward longer 
wavelengths, as more surface area is present in the cooler 
outer regions of the disk.  Increasing $\dot{M}$ tends to shift 
the peak of the accretion disk SED toward shorter wavelengths, 
as the temperature in the inner disk increases.  

Our accretion disk models for V592 Cas were composed of 100 
concentric rings, each of
which contains 1441 azimuthal sections (corresponding to 
0.25$^{\circ}$ per section).  In addition, the outer edge of
the accretion disk also contributes to the total disk SED.
A wavelength- and temperature-dependent correction for 
limb darkening of the accretion disk is applied 
to each azimuthal section
using linear limb darkening coefficients interpolated 
over the grid presented
by \citet{vanhamme93}\footnote{We also tested a wavelength-dependent 
quadratic limb darkening law based on observations of the 
Sun \citep{cox00}, but this did not produce any significant 
difference compared to the linear limb darkening law.  
Limb darkening corrections were not applied to the stellar components.  
In the case of the secondary star, this is because the component 
SED is based on empirical templates which are intrinsically 
limb-darkened.  In the case of gray atmosphere limb darkening, 
which may be most appropriate for the WD (e.g., \citealt{warner71}), 
the limb darkening correction is equivalent to a small 
($\lesssim5$\%; \citealt{davis00}) change in the assumed radius 
of the WD, which is already uncertain by at least this amount.  
In addition, limb darkening is not applied to the circumbinary 
dust disk, since it is treated as optically thin.}.  
In the case of a flat disk, this is a relatively simple 
procedure, as the disk face everywhere presents the same viewing 
angle (i.e., angle between the line-of-sight and a normal to the 
disk face), so 
only the disk edge is limb-darkened across a range of viewing angles.  
However, in the case of a flared
disk (see below), which presents different viewing angles for 
different parts of the disk, limb darkening is calculated 
individually for each
azimuthal section in each ring.  The primary effect of limb 
darkening is to make the short wavelength end of the SED fainter 
relative to the long wavelength end.

By default, the standard model radial 
temperature profile from \citet{FKR} (their Equation 5.41) is 
used to determine the temperature in each disk ring.  
It is possible for the standard model radial temperature 
profile ($T \propto r^{-\gamma}$ with $\gamma = 0.75$) to 
be modified by irradiation; that is, by reabsorption 
in the outer disk of flux from the central object and/or inner disk.  
Irradiation is primarily important in flared disks (i.e., disks 
whose height increases with radius), which can more easily 
intercept luminosity from the WD and inner disk. 
In order to account for this possibility, we have provided an 
option to utilize the irradiation prescription
of \citet{orosz03}, 
in which the radial temperature profile of the outer accretion 
disk (at $R_{\rm acd} \ge R_{\rm crit}$, where 
$R_{\rm crit} = ((7-8\gamma_{\rm out})/(6-8\gamma_{\rm out}))^{2} 
\, R_{\rm WD}$) is allowed to have a shallower temperature 
gradient than in the standard model ($\gamma_{\rm out} < 0.75$). 
At the same time, the height of the disk for radii larger than 
$R_{\rm crit}$ is assumed to increase as $R^{9/8}$ (equation 2.51b 
in \citealt{warner95}).
The outer edge of the disk, which does not ``see'' the WD and 
inner disk, is always assumed to have a non-irradiated temperature 
appropriate for the standard model radial temperature profile.

As found in \citet{orosz03} and 
other studies of the effect of irradiation on model 
accretion disk SEDs (e.g., \citealt{wade88}), we found that irradiation 
(like limb darkening) 
primarily affects the short wavelength end of the SED.  
However, irradiation (which effectively raises the temperature at
a given radius over the non-irradiated case) 
tends to make the short wavelength end of
the SED brighter relative to the long wavelength end, which
somewhat counteracts the effect of limb darkening.
Changing (increasing) the mass transfer rate can 
often compensate for the combined effects of limb darkening and
irradiation in the model accretion disk SED compared to the flat 
disk (i.e., no irradiation) case.  
However, for small values
of $\gamma_{\rm out}$, the model SED becomes
too faint at the short wavelength end, such that increasing $\dot{M}$
enough to supply the ``missing'' UV flux causes the accretion disk
to exceed the observed SED at longer (e.g., IRAC bandpass) wavelengths.
For V592 Cas, this occurs for $\gamma_{\rm out}\lesssim0.60$, 
which requires $\dot{M}\gtrsim1.8\times10^{-8} \, M_{\odot}$ yr$^{-1}$ 
to reproduce the short wavelength end of the SED, but then exceeds 
(with the accretion disk component alone) the observed flux densities 
in the 2MASS and IRAC bands.
Figure \ref{f:sed} shows SED models for V592 Cas using either 
a limb-darkened
flat accretion disk or a limb-darkened flared accretion disk 
with irradiation ($\gamma_{\rm out}=0.7$).  There is no significant 
difference in the resultant total model SEDs.   The corresponding 
model parameters for each case are given in Table \ref{t:model}.  
In the remaining analysis, we utilize the flared accretion disk model.

It is not surprising that the accretion disk of a nearly 
face-on novalike CV is very bright.  Indeed, the 
nominal accretion disk component for V592 Cas dominates 
at all wavelengths from the optical through near-IR, and 
even into the mid-IR (IRAC) range, accounting for all but 
$\lesssim10$\% of the total flux density in the wavelength 
range 0.5--5 $\mu$m\footnote{Many CVs are inferred to have 
a bright spot on the edge of the accretion disk at the site 
of the initial accretion stream impact.  In principle, this 
could introduce another component into the model.  
We experimented with representing the bright spot in our 
V592 Cas SED model by using an additional blackbody component 
with temperature of 20,000 K (i.e., peaking in the 
near-UV) and projected emitting area up to several 
times the projected area of the WD.  However, the result is 
a contribution of only $<1$\% of the total flux density at 
all wavelengths, and $<0.5$\% at all wavelengths longer than 
1 $\mu$m. Since the flux density uncertainties at the blue 
end of the SED range from 10--20\% from the optical to UV 
(mainly due to the dereddening correction), the contribution 
of a disk bright spot is below the level of detectability.  
In a higher inclination CV, in which the relative contribution 
from the acccretion disk face is reduced, a bright spot 
component could be an important addition to the SED model.
Additional components, such as a boundary layer, optically 
thin disk corona or wind, etc., could also contribute to the 
accretion disk SED, but are not considered here because they 
would primarily affect only the UV end of the spectrum, have 
relatively sparse empirical and theoretical bases in the mid-IR, 
and would exceed the ability of the current data set to constrain.}.  
This makes accounting for the flux 
density that is in excess of the accretion disk model 
component at long wavelengths ($\lambda\gtrsim8$ $\mu$m) 
the principle challenge to reproducing the observed SED 
of V592 Cas. 

\begin{figure}
\epsscale{0.95}
\plotone{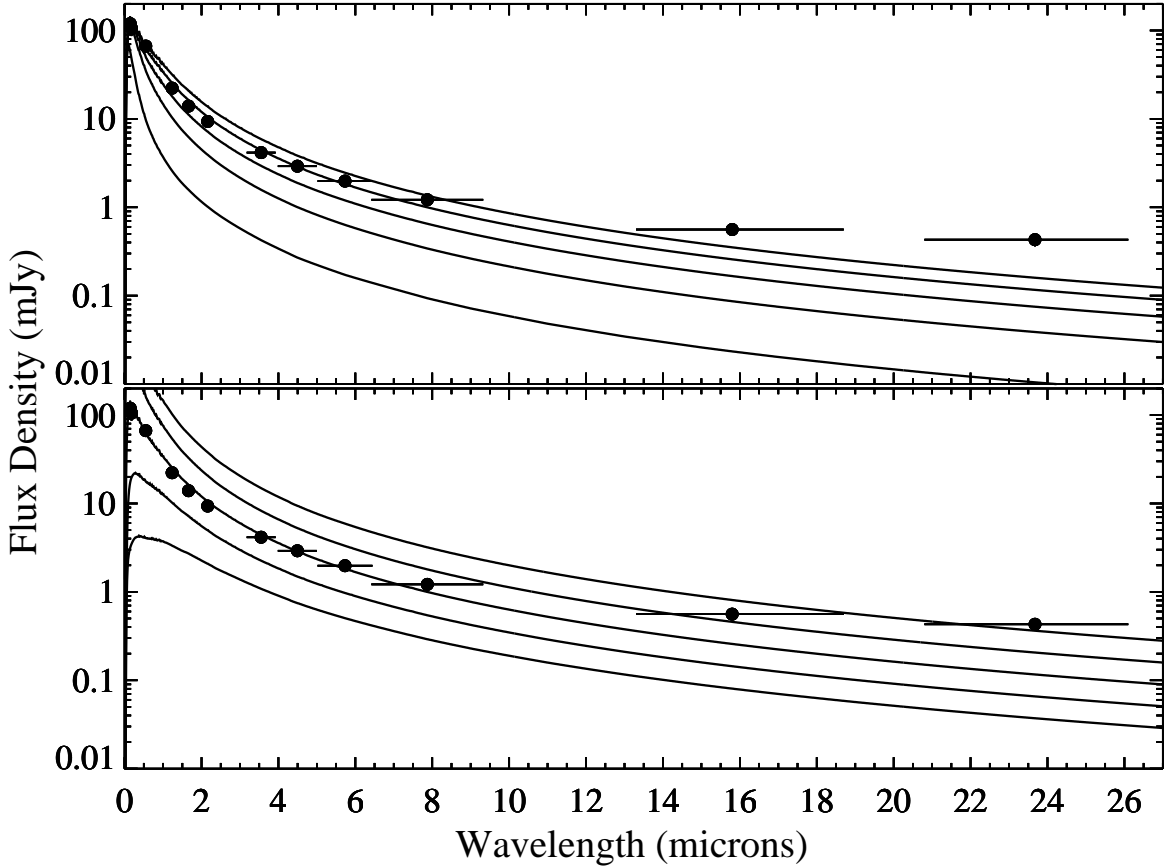}
\epsscale{1.00}
\caption{As in Figure \ref{f:sed}, but showing a model that 
does not contain a circumbinary dust disk, in which a 
steady state accretion disk is the dominant component 
at all wavelengths.  
For clarity, only the total system model is plotted; 
the WD and secondary star are as shown in Figure \ref{f:sed}.
The top panel shows the effect of changing the outer radius 
of the accretion disk; from bottom to top the model curves 
have $R_{\rm acd,out} = 5, 15, 25, 35, 45 R_{\rm wd}$ 
(the nominal value is $35 R_{\rm wd}$).  The bottom panel shows 
the effect of changing the mass transfer rate through the 
accretion disk; from bottom to top, the model curves 
have $\dot{M} = 0.01, 0.1, 1, 10, 100$ times the nominal value 
of $1.5\times10^{-8} \, M_{\odot}$ yr$^{-1}$.  
All other model parameters are as given in Table \ref{t:model}.  
\label{f:sedgrid}}
\end{figure}

Figure \ref{f:sedgrid} shows the effect of changing 
$R_{\rm acd,out}$ (top panel) or $\dot{M}$ (bottom panel) in the 
nominal flared accretion disk component.  It is clear from this 
figure that adjusting either parameter individually can not 
account for the flux density excess at long wavelengths.  
In the case of $R_{\rm acd,out}$, the maximum allowed value 
is $\approx45 R_{\rm wd}$, corresponding to the size of the 
WD Roche lobe, and even this is insufficient to reproduce 
the 16- and 24-$\mu$m measurements.
Increasing $\dot{M}$ also fails -- when the model is 
bright enough to match the 
16-$\mu$m measurement, then it is not bright enough to match at 
24 $\mu$m, and when it would be bright enough to match the 24-$\mu$m 
measurement, then it would be too bright at 16 $\mu$m.  
The latter case would require an exceptionally high mass 
transfer rate, in excess of $10^{-6} \, M_{\odot}$ yr$^{-1}$, 
which is not plausible. 
Even worse, in order to match even the 16-$\mu$m measurement by 
increasing either parameter (even disregarding the firm upper 
limit to $R_{\rm acd,out}$), the model flux 
density greatly exceeds the observed values at shorter 
wavelengths.  Changing both parameters in unison (e.g., 
increasing $R_{\rm acd,out}$ while decreasing $\dot{M}$) also 
does not produce a viable solution, since this does not change 
the slope of the long wavelength end of the accretion disk 
model component -- it will never match both the 16- and 
24-$\mu$m measurements at the same time, and will always be too 
bright at shorter wavelengths when it is forced to match 
either of the longer wavelength measurements.  We are left with 
the conclusion that, in the context of the standard steady 
state accretion disk model (with or without irradiation), 
an additional component must be 
present to explain the observed 16- and 24-$\mu$m data -- 
this is discussed in \S\ref{s:cbd}.

The nominal accretion disk outer radius is 
$R_{\rm acd,out}=35 R_{\rm wd}$, which is equivalent to 
about 75\% of the WD Roche lobe radius.  This is consistent 
with estimates for the sizes of accretion disks in novalike 
CVs, which, at the mass ratio of V592 Cas, range from 
$\sim0.3 R_{\rm L1}$ from the theory for a zero viscosity 
disk to $\sim0.4$--$0.8 R_{\rm L1}$ from disk reconstructions 
based on eclipse observations \citep{ritter80,sulkanen81,rutten92}.  
The mass transfer rate, $\dot{M}\approx1\times10^{-8} \, M_{\odot}$ yr$^{-1}$ 
is in the middle of the $\sim10^{-9}$--$10^{-8} \, M_{\odot}$ yr$^{-1}$ 
range that is often cited as characteristic of novalike CVs 
\citep{warner95}.  \citet{taylor98}
estimated a mass transfer rate of $9\times10^{-9} \, M_{\odot}$ yr$^{-1}$ 
for V592 Cas from a power law fit to an optical--near-IR SED, 
which is quite close to our result (we note that if we utilize a 
flat standard model accretion disk with no irradiation and no limb 
darkening, then our best match to the observed SED is obtained with 
$\dot{M}=9\times10^{-9} \, M_{\odot}$ yr$^{-1}$, as found 
by \citealt{taylor98}).

The orbital period of V592 Cas places 
it at the upper end, or well inside, the CV orbital period 
gap (depending on whose period limits for the gap are used; 
e.g., 2.1--2.8 hr, \citealt{patterson84}; 
2.10--2.85 hr, \citealt{howell01}, 2.19--2.75 hr, 
\citealt{shafter92}; 2.15--3.18 hr, \citealt{knigge06}; 
2.25--2.75 hr, \citealt{politano07}; 2.3--2.8 hr, \citealt{warner95}).  
From extrapolating between the expected mass transfer rates at 
either end of the period gap (e.g., \citealt{warner95,howell01}), 
we would expect that at the orbital period of V592 Cas, 
in the absence of whatever mechanism causes the gap, it should 
have $\dot{M}\sim10^{-9} \, M_{\odot}$ yr$^{-1}$, an 
order of magnitude smaller than inferred here and in 
\citet{taylor98}.  However, the observed spread in $\dot{M}$ at 
the long period end of the gap ranges from 
$\sim10^{-10}$--$10^{-8} \, M_{\odot}$ yr$^{-1}$ 
\citep{patterson84,kolb01}.
In any case, the mere fact that V592 Cas is a CV inside the 
period gap but still accreting at a high rate implies that it 
is not a run of the mill CV.  It has either recently evolved 
into the gap and is in the process of shutting down mass 
transfer, or is a relatively young CV that only recently 
initiated mass transfer at close to its current orbital 
period\footnote{A third possibility is that V592 Cas is an 
old CV which is evolving to longer orbital period out of the 
gap due to the evolutionary effect of a massive circumbinary 
disk (as discussed in \citealt{willems07}); however, in light 
of the low dust mass that we have found -- see \S\ref{s:cbd} 
and \S\ref{s:totalmass}--\ref{s:thick} -- we consider this 
scenario unlikely.}.

\subsubsection {The Circumbinary Dust Disk}
\label{s:cbd}

In order to supply the requisite additional flux density 
at the longest mid-IR wavelengths, we utilize a circumbinary 
dust disk component.
This model component is calculated as described in 
\citet{hoard07}, under the assumption that the disk is 
optically thin, and composed of spherical grains with 
radius of 1 $\mu$m that (re)radiate as blackbodies.  
The problem of determining the inner boundary condition of 
temperature in the presence of multiple irradiating bodies 
(i.e., the WD, secondary star, and accretion disk), as 
described in \citet{brinkworth07}, is also an issue for 
V592 Cas.  Because the WD in V592 Cas is relatively hot, 
and both the secondary star and accretion disk are large, 
we expect that the maximum radius at which the dust temperature 
would be above 1000 K (i.e., in the 1000--2000 K range for 
dust sublimation) is on the order of 200--300 $R_{\rm wd}$ 
in V592 Cas\footnote{This also likely rules out a substantial circumstellar 
dust ring around the WD -- as discussed for the short 
orbital period dwarf nova WZ Sge in \citet{howell08} -- 
as the source of the long wavelength IR excess in V592 Cas.}.  
This is well beyond the tidal truncation limit of $\approx1.5$ 
times the binary separation (about $157 R_{\rm wd}$ for 
V592 Cas), which we have assumed for the inner edge radius 
of the circumbinary disk in previous works.  
Consequently, we have used 
the ``equivalent star'' prescription given in 
\citet{brinkworth07} to set the temperature at the inner 
edge of the circumbinary disk for a given radius.  
We note that essentially identical circumbinary disk SEDs can be obtained 
using different combinations of inner and outer radii by 
adjusting the total mass in the disk (e.g., we can obtain 
an SED identical to the one shown in Figure \ref{f:sed} 
by forcing the circumbinary disk inner edge radius to be at the tidal 
truncation radius, if the total mass of dust in the disk 
is increased by a factor of $\approx5$).  
Similarly, the effect of different prescriptions for the 
radial temperature profile for the dust can be offset by 
adjusting the inner/outer radii and/or total mass 
of the circumbinary disk.
In the context of reproducing the observed data, the 
important parameter is the temperature at the inner edge 
of the disk -- it cannot be significantly higher than 
$\sim500$ K, or the circumbinary disk component SED becomes too bright 
at short wavelengths to match the observations.

\section{Discussion}

\subsection{Excluding Bremsstrahlung Emission}

\citet{howell06} discuss several general reasons why 
bremsstrahlung, or free-free, emission in an outflowing 
wind can be excluded as a significant source of mid-IR emission in CVs.
\citet{dubus04} present quantitative observational tests that would 
confirm or refute IR emission in CVs as being caused by bremsstrahlung 
emission.  The first of these relates to the expected 
$f_{\nu}\propto\nu^{0.6}$ spectrum of bremsstrahlung emission.  
At 16 $\mu$m, the IR excess in V592 Cas (i.e., the remaining 
flux after the wd+ss+acd model components have been subtracted) 
is 0.33 mJy; at 24 $\mu$m, the excess is 0.30 mJy.  
This is a ratio of $f_{\nu,24}/f_{\nu,16}=0.91$, whereas 
if the IR excess followed a spectral law consistent with 
bremsstrahlung emission, then we should expect the flux 
density at 24 $\mu$m to be 
$\approx0.78\times$ the flux density at 16 $\mu$m.

The second test relates to the wind mass loss rate required to 
explain the observed level of IR excess via bremsstrahlung 
emission.  Using Equation 4 from \citet{dubus04} (which is 
based on the work of \citealt{wright75}) with an assumed wind 
velocity of 2000 km s$^{-1}$ (\citealt{kafka08} found wind 
velocities reaching up to 5000 km s$^{-1}$ in V592 Cas), the 
observed IR excesses at 16 and 24 $\mu$m would require wind 
mass loss rates of $\approx4.0$--$5.2\times10^{-9} \, M_{\odot}$ yr$^{-1}$.  
This is equivalent to $\approx$25--35\% of the 
entire inferred mass transfer rate 
from the secondary star in V592 Cas and, as such, is improbably 
high for the wind mass loss rate.

\subsection{No Dust Emission at Short Wavelengths}

\citet{belle04} concluded that there is no evidence for 
the presence of a circumbinary dust disk in the IR SED 
for V592 Cas; however, their data spanned only the 
$JHKL^{\prime}$ bands (i.e., $\approx1.2$--$3.8$ $\mu$m).  
From our data, it is clear that the IR excess in V592 Cas 
is not readily apparent until well past 5 $\mu$m.  
For example, at wavelengths $\lesssim5$ $\mu$m, the 
wd+ss+acd components account for $>90$\% of the observed 
flux density, whereas at wavelengths $\gtrsim8$ $\mu$m, 
these components account for $<70$\% of the observed 
flux density.  At $L^{\prime}$, only 3\% of the observed 
flux density is not accounted for by the wd+ss+acd components.  
In addition, \citet{belle04} were looking for a much 
hotter circumbinary disk component, with $T\sim3000$ K, whereas our 
observations show that the maximum temperature of the 
dust is $T\approx500$ K.
We want to emphasize that this is {\em not} just a case in 
which a bright, near-face-on accretion disk is ``masking'' 
fainter dust emission at short wavelengths.  Dust at hotter 
temperatures in V592 Cas would have been detectable at 
$\lambda\lesssim5$ $\mu$m, even against the comparatively 
bright contribution of the accretion disk at those wavelengths, 
if the long wavelength end of the dust distribution was bright
 enough to reproduce the observed 16- and 24-$\mu$m measurements.

\subsection{Dissipation and Formation of the Dust}
\label{s:dustform}

One question that has plagued a full understanding of the 
presence of circumbinary disks in CVs is the origin of the dust.  
Possible scenarios include:\ 
a remnant of the common envelope phase of the CV's early evolution, 
formation through tidal disruption of minor planets or comets 
that survived from the progenitor solar system (as assumed for 
the origin of circumstellar dust disks around single WDs), 
creation via mass outflows in winds and nova outbursts from the 
inner binary (the mechanism assumed for the formation of 
circumbinary dust disks in the theoretical models of 
\citealt{willems07}), or some combination of these processes.
Regardless of the exact origin of the dust, since the calculation 
of the structure of 
the circumbinary disk is predicated on the assumption that there is a 
region in the CV system in which temperatures are too high 
for dust to exist, it seems likely that the circumbinary disks cannot be 
entirely static:\ dust will be sublimated at some rate at 
the inner edge as grains stray too close to the inner 
binary, and another trickle of dust will be lost at the 
outer edge of the circumbinary disk as grains stray too far away from 
the gravitational influence of the inner binary.
So, it might be more appropriate to wonder about not only the 
origin of the dust, but its longevity as well.
In V592 Cas, at least, there is an obvious mechanism to 
provide a continuing supply of raw materials for the circumbinary disk, 
namely, the wind outflow discussed in \citet{witherick03},
\citet{prinja04}, and \citet{kafka08}.  

The total circumbinary dust disk mass in V592 Cas amounts to only 
$\sim10^{-12} \, M_{\odot}$, which is a small fraction 
($\sim0.01$\%) of the total annual mass transfer budget 
in this system.  If the total mass flux present in the 
wind outflow from V592 Cas is a 
tenth of a percent of the total mass 
transfer budget, and even if the efficiency of capturing 
material from the wind into the circumbinary disk is a similarly small 
fraction, then this mechanism can still completely 
replenish the circumbinary disk on a timescale of $\sim100$ yr.
Another way to look at this is that as long as the mass loss
rate from the circumbinary disk (e.g., dissipation through 
sublimation and/or escape from the system)
is smaller than the rate of dust formation from the wind, then the
wind can maintain the circumbinary disk at a constant mass.  
Using the 0.1\% efficiencies for the wind outflow and 
dust formation rates, as suggested above, 
the circumbinary disk mass
will be maintained as long as $\lesssim1$\% of the total
circumbinary disk mass is lost per year.  
This result also implies that in V592 Cas a much more 
massive circumbinary disk probably could not be maintained 
by the wind outflow alone.

\subsection{Total Mass of Circumbinary Dust}
\label{s:totalmass}

The total mass in our model circumbinary dust disk is 
$M_{\rm cbd}\approx10^{21}$ g.  Reasonable variations in 
the inner radius of the circumbinary disk that would be required to 
yield the necessary boundary condition of 
$T_{\rm cbd,in}\approx500$ K in the presence of different 
irradiation prescriptions (i.e., treatment of multiple 
heating sources in the inner binary) can be balanced 
against changing the total mass in the circumbinary dust 
disk to produce an SED that is 
indistinguishable from the one presented here.  
In addition, the photometric uncertainties also provide some 
leeway for changing the dust mass while still adequately 
reproducing the observed SED.
However, these variations in mass are relatively small 
(i.e., factor of $\lesssim5$).  The mass of the circumbinary disk in 
V592 Cas exceeds those found in several of our past 
investigations of short orbital period magnetic CVs, 
which were on the order of $10^{15}$--$10^{17}$ g 
\citep{brinkworth07}.  It is comparable to the high end 
of the range of mass estimates for the circumbinary disk in the short 
orbital period, magnetic CV EF Eri \citep{hoard07}.  
However, as found in all of these previous studies, 
the total mass of dust is many orders of magnitude 
smaller than the $10^{28}$--$10^{29}$ g predicted to be 
necessary for circumbinary disks to serve as angular momentum loss 
mechanisms that can significantly affect the secular 
evolution of CVs \citep{taam03,willems07}.  
The prediction by \citet{willems07} that the circumbinary 
dust disks in (disk-accreting) CVs should only be detectable 
at wavelengths longer than 10 $\mu$m is borne out by our observations 
for the case of V592 Cas.

The outer radius of the circumbinary disk has been somewhat arbitrarily 
set at 15,000 $R_{\rm wd}$, giving an outer temperature of 
50 K, comparable to what one might expect as the disk merges 
with the local ISM.  The model SED at the observed wavelengths 
is not sensitive to decreasing this outer radius by more than 
a factor of 2, corresponding to an outer temperature of 
$\sim100$ K.  However, in order to continue to match the 
observed mid-IR photometry, the total mass of dust in the 
disk must be decreased as the outer radius decreases -- this 
exacerbates the low mass discrepancy of the circumbinary disks 
compared to the expectation from CV evolution theory.  Unfortunately, 
applying the reverse of this scenario also cannot resolve 
the low mass problem:\ increasing the outer radius of the 
circumbinary disk to as much as 100,000 $R_{\rm wd}$ (corresponding to a 
rather low temperature of only 12 K) only increases the 
requisite total dust mass by a factor of $\sim7$.

\subsection{Excluding an Optically Thick Circumbinary Disk}
\label{s:thick}

The evolutionary calculations of \citet{taam03} and \citet{willems07} 
assume that the circumbinary disks are massive ($\gtrsim10^{28}$ g), 
hot ($T_{\rm cbd,in}\approx1500$--$3000$ K), and optically thick.  
If the circumbinary disks in CVs do not have these characteristics, 
then we should not expect them to act as effective angular momentum 
loss mechanisms during the course of CV evolution.
\citet{brinkworth07} and \citet{hoard07} discussed the reasons for 
preferring an optically thin circumbinary disk model over an 
optically thick one when attempting to match the 
observed IR excesses of magnetic CVs.
For the sake of completeness, and because it would offer the 
simplest method for substantially increasing the dust mass in 
V592 Cas compared to the optically thin case, we have also 
explored the optically thick circumbinary disk case.
We can produce a model circumbinary disk 
SED very similar to that shown in Figure \ref{f:sed} using an 
optically thick treatment.  However, we must force the 
circumbinary disk to have an inner edge temperature of 
$T_{\rm cbd,in}\approx500$ K at the tidal truncation 
radius -- this is a factor of $\approx3$--4 cooler than the 
expected temperature at this radius due to irradiation from 
the WD.  Starting the circumbinary disk at a larger radius 
where the expected temperature is closer to $\approx500$ K 
does not work because the correspondingly larger surface area 
of the circumbinary disk makes the SED component too bright 
to match the observed mid-IR data.  In addition, the 
circumbinary disk must be truncated at an outer radius 
corresponding to a temperature of $\approx250$ K, much warmer 
than the ambient temperature of the ISM, or the model SED 
component is too bright at the long wavelength end of 
the observed range.
These inconsistencies again suggest that the optically 
thick case is not appropriate for matching the shape of the 
observed IR excess in V592 Cas.  Thus, we are led to conclude 
that the results from the optically thin case are more likely 
to be representative of the distribution of dust.

\subsection{An Initially Massive Circumbinary Disk?}
\label{s:dustmass}

Considering the low dust masses that we have found in several 
CVs (now including a high $\dot{M}$ novalike CV with a wind; 
that is, a good candidate to have a large amount of circumbinary 
material), we wonder if an initially massive circumbinary disk 
(perhaps formed during the common envelope phase of the CV's 
early history) could evolve over time into a low mass remnant 
that we see now.
To begin to address this question, we have used a ``toy'' model 
to explore the general behavior of an initially massive 
circumbinary dust disk in the presence of dissipation at the 
disk edges and contribution from formation of new dust using 
material in the wind from the inner binary, as described 
in \S\ref{s:dustform}.  This model is purely parametric, and 
does not consider the precise physics of how a particular rate 
of dust dissipation or formation might arise.
The total mass of circumbinary dust as a function of time (years) 
is simply calculated under the assumption that a constant small 
fraction ($\varepsilon_{1}$) of the total dust mass is lost 
each year through dissipation.  At the same time, a small 
fraction ($\varepsilon_{2}$) of the mass transfer rate in 
the CV ($\dot{M}$) is carried away in a wind, of which 
another small fraction ($\varepsilon_{3}$) forms dust and 
contributes to the mass of the circumbinary disk.  That is,
\begin{equation}
M_{\rm cbd,i} = (1-\varepsilon_{1})M_{\rm cbd,i-1} + \varepsilon_{2}\varepsilon_{3}\dot{M},
\label{e:dustmass}
\end{equation}
and this equation is evaluated over $i=1$--$10^{8}$ yr, 
with initial mass of at least $10^{28}$ g at $i=0$.  
We assume that $\log(\dot{M})\propto -\log(i)$ (i.e., the 
mass transfer rate decreases linearly in log-log space), 
with an initial value of 
$\dot{M}_{\rm initial}=10^{-7.5} \, M_{\odot}$ yr$^{-1}$, 
and evolve the system over $10^{8}$ yr to a final mass 
transfer rate of $10^{-8} \, M_{\odot}$ yr$^{-1}$ or smaller.  
This evolving mass transfer rate, to first order, mimics 
the expected decrease in $\dot{M}$ during the evolution of 
a CV above the period gap (e.g., \citealt{shafter92,warner95,howell01}).  

\begin{figure}
\epsscale{0.78}
\plotone{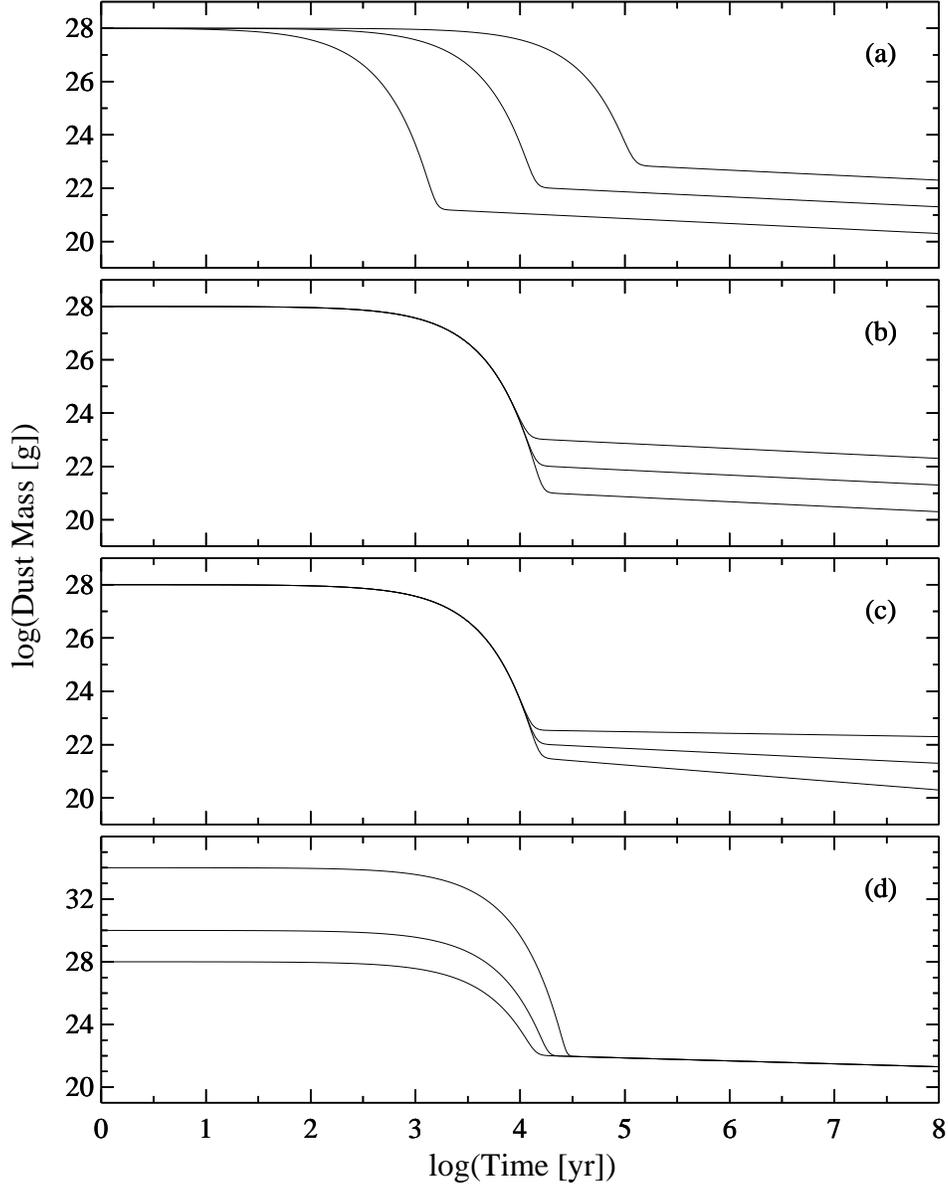}
\epsscale{1.00}
\caption{Evolution of circumbinary dust mass as calculated using 
the toy model described in \S\ref{s:dustmass}.  
In panels (a), (b), and (d), the CV's mass transfer rate evolves from 
$10^{-7.5} \, M_{\odot}$ yr$^{-1}$ to 
$10^{-9} \, M_{\odot}$ yr$^{-1}$ over $10^{8}$ yr.
Panel (a) shows the effect of changing the dust dissipation 
efficiency, $\varepsilon_{1}$, with values of 0.01, 0.001, and 
0.0001 (curves from bottom to top, respectively) while keeping 
the dust contribution efficiencies fixed at 
$\varepsilon_{2}=\varepsilon_{3}=0.001$.  
Panel (b) shows the effect of changing the product of 
the dust formation efficiencies, $\varepsilon_{2}\varepsilon_{3}$, 
with values of 0.0001, 0.00001, and 0.000001 (curves from top 
to bottom, respectively) while keeping the dust dissipation 
efficiency fixed at $\varepsilon_{1}=0.001$.  
In panels (c) and (d)
the dust dissipation and formation efficiencies were fixed 
at $\varepsilon_{1}=\varepsilon{2}=\varepsilon{3}=0.001$.
Panel (c) shows the effect of changing the final mass transfer 
rate, with values of $10^{-8}$, $10^{-9}$, and 
$10^{-10} \, M_{\odot}$ yr$^{-1}$ (curves from top to bottom).
Panel (d) shows the effect of changing the initial mass of 
circumbinary dust, with values of $10^{28}$, $10^{30}$, and 
$10^{34}$ g (curves from bottom to top).
Note the different vertical axis scale in panel (d).
\label{f:dustmass}}
\end{figure}

Several examples of the outcome of this calculation are shown 
in Figure \ref{f:dustmass}.  In general, the dust mass follows 
a similar evolution in all cases, over a wide range of values 
for $\dot{M}_{\rm final}$, $M_{\rm cbd, i=0}$, $\varepsilon_{1}$, 
$\varepsilon_{2}$, and $\varepsilon_{3}$.  
It first goes through a period of nearly exponential decline 
during which the rate of mass contributed to the circumbinary 
dust as new dust forms is insignificantly small compared to 
the total mass of dust, and the losses from the circumbinary 
dust (parameterized by $\varepsilon_{1}$) dominate.  At some 
critical time, $\tau_{\rm crit}$ (marked by a sharp ``elbow'' 
in the plot of $\log({\rm time})$ vs.\ $\log({\rm dust \, mass})$), 
the rate of mass being 
contributed to the circumbinary dust by the wind (parameterized 
by $\varepsilon_{2}$ and $\varepsilon_{3}$) becomes comparable 
to the rate of mass being lost, and the total dust mass becomes 
relatively stable for the remaining evolution of the system 
(showing at most only a gradual decline by less than 
an order of magnitude over many tens of millions of years). 
In general, the value 
of $\tau_{\rm crit}$ and the final dust mass both increase as 
$\varepsilon_{1}$ is decreased because mass is lost from the 
dust at a lower rate and, consequently, it takes longer for 
the contribution from the wind to become significant in 
comparison to the rate of dust dissipation and the total mass of dust.  
The value of $\tau_{\rm crit}$ also increases, to a lesser extent, 
when the initial dust mass is larger, because there is more dust 
to dissipate before the contribution from the wind becomes 
significant; however, if all other parameters are fixed, the 
evolution of the dust mass after $\tau_{\rm crit}$ is identical 
regardless of the value of $M_{\rm cbd, i=0}$.

Because $\varepsilon_{2}$ and $\varepsilon_{3}$ are used as a 
product in Equation \ref{e:dustmass}, it does not matter for 
the sake of calculation which parameter is changed; however, 
they do have different physical interpretations, as defined above.  
Changing the product of $\varepsilon_{2}$ and $\varepsilon_{3}$ 
has only a small effect on $\tau_{\rm crit}$ for a given 
$\varepsilon_{1}$.  This mainly affects the final total mass 
of dust, with the mass increasing as the product 
$\varepsilon_{2}\varepsilon_{3}$ increases, which causes the 
rate of dust formation out of the CV wind to increase and, 
consequently, become significant earlier compared to the 
total dust mass and rate of dust dissipation.

Panel (a) in Figure \ref{f:dustmass} shows the result 
of varying the circumbinary dust dissipation rate 
efficiency, $\varepsilon_{1}=0.01, 0.001, 0.0001$, 
with the wind-fueled 
dust formation rate efficiencies fixed at
$\varepsilon_{2}=\varepsilon_{3}=0.001$.
The result of this is to cause $\tau_{\rm crit}$ to vary 
from $\sim10^{3}$--$10^{5}$ yr and the final dust mass to 
vary from $\sim2\times10^{20}$--$2\times10^{22}$ g as 
$\varepsilon_{1}$ decreases.  By comparison with the other
cases (discussed below), it is clear that the value
of $\varepsilon_{1}$ has the largest effect on $\tau_{\rm crit}$.

Panel (b) in Figure \ref{f:dustmass} shows the result 
of varying the product of the 
wind-fueled dust formation rate efficiencies,
$\varepsilon_{2}\varepsilon_{3}=0.0001, 0.00001, 0.000001$, with the
circumbinary dust dissipation rate 
efficiency fixed at $\varepsilon_{1}=0.001$.
In this scenario, the value of $\tau_{\rm crit}$ is relatively 
unchanged at $\sim10^{4}$ yr, but the final dust mass varies 
from $\sim2\times10^{20}$--$2\times10^{22}$ g as 
$\varepsilon_{2}\varepsilon_{3}$ increases.
In both panels (a) and (b) the initial dust mass and final mass 
transfer rate were fixed at $M_{\rm cbd, i=0}=10^{28}$ g and 
$\dot{M}_{\rm final}=10^{-9} \, M_{\odot}$ yr$^{-1}$, respectively.

Panel (c) in Figure \ref{f:dustmass} shows the result 
of varying the final mass transfer rate, 
$\dot{M}_{\rm final}=10^{-8}, 10^{-9}, 10^{-10} \, M_{\odot}$ yr$^{-1}$, 
with the dust dissipation and formation efficiences fixed at 
$\varepsilon_{1}=\varepsilon_{2}=\varepsilon_{3}=0.001$.
For the dust mass evolution up to $\tau_{\rm crit}$, 
the result is very similar to the case shown in panel (b).
After $\tau_{\rm crit}$, 
the total dust mass declines more slowly for larger values of 
$\dot{M}_{\rm final}$, since there is a correspondingly larger 
wind outflow from which new dust can form.  The case for 
$\dot{M}_{\rm final}=10^{-8} \, M_{\odot}$ yr$^{-1}$ corresponds 
approximately to V592 Cas at the current time, and shows the least 
change in total dust mass in the post-$\tau_{\rm crit}$ evolution 
of the CV.  The case for 
$\dot{M}_{\rm final}=10^{-9} \, M_{\odot}$ yr$^{-1}$ might be 
considered more ``typical'' of CVs at this orbital period,
or could also represent the future of V592 Cas after continued evolution
further reduces its mass transfer rate.

Finally, panel (d) in Figure \ref{f:dustmass} shows the result 
of changing the initial dust mass, 
$M_{\rm cbd, i=0}=10^{28}, 10^{30}, 10^{34}$ g, 
with the dust dissipation and formation efficiences fixed at 
$\varepsilon_{1}=\varepsilon_{2}=\varepsilon_{3}=0.001$ and the 
final mass transfer rate fixed at 
$\dot{M}_{\rm final}=10^{-9} \, M_{\odot}$ yr$^{-1}$.  
We had initially chosen $10^{28}$ g of dust as a lower limit 
starting point of interest; obviously, the case in which the 
initial mass of dust is already smaller than predicted to be 
necessary to affect CV evolution is of little interest.  
However, if the initial dust originates, for example, during the 
common envelope phase, then there could potentially be more than 
this lower limit case.  However, as panel (d) shows, even an 
initial mass as much as $10^{6}$ times larger than the lower 
limit results in identical 
post-$\tau_{\rm crit}$ conditions and evolution, with only a small 
increase in $\tau_{\rm crit}$ as $M_{\rm cbd, i=0}$ increases.

In general, we find that this simple toy model suggests that 
even an initially very massive ($>>10^{28}$ g) circumbinary 
dust disk present at the birth of a CV will rapidly evolve 
(in only $10^{3}$--$10^{5}$ yr) into a disk with much lower 
mass ($\sim10^{20}$--$10^{22}$ g).  The dust disk mass is 
then approximately stable over the remaining evolution of 
the CV. The only requirements to produce this scenario are 
that the circumbinary dust experiences ongoing low rates of 
dissipation (e.g., via sublimation and escape from the 
system's gravity) and formation (e.g., from material carried 
outward in a wind from the inner binary).  If a massive 
circumbinary dust disk is a common by-product of the common 
envelope phase of CV formation, then such a scenario might 
explain why all of the CVs so far observed with {\em Spitzer} 
appear to have only small amounts of circumbinary dust.  
We are encouraged by the relatively close agreement between 
our inferred mass of dust in V592 Cas and the toy model 
prediction for the final mass of circumbinary dust in a 
high mass transfer rate CV that has evolved to the long 
period end of the period gap.

To first order, the initial rapid decrease in the mass of 
circumbinary dust would seem to imply that circumbinary dust 
disks do not play a role in the secular evolution of the 
CV population.  However, this begs a new question:\ 
is the presence of a massive circumbinary disk for a short 
time early in the evolutionary history of a CV sufficient 
to have a lasting impact on the secular evolution of the CV?  
That is, could such a scenario result in an early episode 
of extra angular momentum loss of sufficient magnitude to 
alter the characteristics at the far end of the evolutionary 
sequence of the CV population compared to predictions 
utilizing angular momentum loss via magnetic braking and 
gravitational radiation alone?  The answer to that question 
is beyond the scope of this paper.

\section{Conclusions}

We have found compelling observational evidence for the 
existence of IR excess in the near-face-on, novalike CV V592 Cas.  
This result stems primarily from our new {\em Spitzer} IRS-PUI 
and MIPS observations at 16 and 24 $\mu$m, respectively.  
We have provided a representative model that reproduces the 
observed (dereddened) SED using a combination of white
dwarf, secondary star, steady state accretion disk, 
and circumbinary dust disk components.
At short wavelengths ($\lambda\lesssim5$ $\mu$m), 
the SED of V592 Cas is dominated 
by the steady state accretion disk and, as found in past 
studies of this system, there is no indication of the presence of 
circumbinary material until the SED at longer wavelengths 
is considered.  The total mass in the circumbinary dust disk of 
V592 Cas, consistent with recent studies of IR excess in 
magnetic CVs, is apparently too small to have a significant 
effect on the secular evolution of the CV.  This is the 
first estimate for the mass of circumbinary dust in a non-magnetic, 
disk-accreting CV at an orbital period ($\approx166$ min) 
significantly longer than the observed period 
minimum ($\approx80$ min), and 
demonstrates that the low dust mass results from our 
previously studied systems are not solely a result of the 
fact that they contain strongly magnetic WDs.  

The problem of accounting for the origin and/or longevity of 
the circumbinary dust is mitigated in the case of V592 Cas 
by the presence of a wind outflow, which can easily provide 
the necessary raw materials to replenish the circumbinary disk 
on short timescales or maintain it at a relatively constant mass.  
An implication of this result is that CVs without significant 
wind outflows might be expected to contain little or no 
circumbinary dust.  However, our previous {\em Spitzer} 
observations of magnetic CVs show that dust {\em can} be 
present even in systems that lack a strong wind.  
So, there must be more than one route to the presence of 
circumbinary dust in CVs.

We have made a simple calculation that shows that an initially 
massive circumbinary dust disk, perhaps formed during the 
common envelope phase of the binary's early evolution, will 
rapidly evolve to a relatively stable low mass in the presence 
of ongoing low rates of dissipation of dust in the existing 
circumbinary material and formation of new dust (for example, 
fueled by matter carried outward in a wind from the inner 
binary).  It remains to be demonstrated in detail whether or 
not such evolving circumbinary disks can still play a role 
in the secular evolution of the CV population.

\acknowledgments

This work is based in part on observations made with the 
{\em Spitzer Space Telescope}, which is operated by the Jet 
Propulsion Laboratory, California Institute of Technology, 
under a contract with the National Aeronautics and Space 
Administration (NASA).
Support for this work was provided by NASA.
This work makes use of data products from the 
Two Micron All Sky Survey, which is a joint project of the 
University of Massachusetts and the Infrared Processing and 
Analysis Center/Caltech, funded by NASA and the NSF. 
This research has made use of the SIMBAD database,
operated at CDS, Strasbourg, France.
We acknowledge with thanks the variable star observations 
from the AAVSO International Database contributed by 
observers worldwide and used in this research.


\clearpage

\begin{deluxetable}{lllll}
\tablewidth{0pt}
\tabletypesize{\footnotesize}
\tablecaption{Spectral energy distribution data \label{t:data}} 
\tablehead{
\colhead{ } & 
\colhead{ } & 
\multicolumn{2}{c}{Flux Density:} &
\colhead{ } \\
\colhead{Band} & 
\colhead{Wavelength} & 
\colhead{Observed} &
\colhead{Dereddened\tablenotemark{a}} &
\colhead{Date Obtained} \\
\colhead{ } & 
\colhead{(microns)} & 
\colhead{(mJy)} &
\colhead{(mJy)} &
\colhead{(UT)} 
}
\startdata
\vspace*{4pt}
UV                & $0.1450$                   & $19.99\pm2.00$          & $120\pm25$   & 05 Dec 1981\tablenotemark{b} \\
\vspace*{4pt}
UV                & $0.1800$                   & $20.53\pm2.05$          & $105\pm19$   & 05 Dec 1981\tablenotemark{b} \\
\vspace*{4pt}
V                 & $0.55\pm0.045$             & $36.11^{+3.36}_{-3.65}$ & $66.6\pm7.0$ & 08 Oct 2005 \\
\vspace*{4pt}
2MASS-J           & $1.235^{+0.125}_{-0.115}$  & $19.27\pm0.53$          & $22.28\pm1.85$ & 28 Sep 2000 \\
\vspace*{4pt}
2MASS-H           & $1.662^{+0.118}_{-0.152}$  & $12.82^{+0.44}_{-0.45}$ & $13.95\pm1.32$ & 28 Sep 2000 \\
\vspace*{4pt}
2MASS-K$_{\rm s}$ & $2.159^{+0.141}_{-0.139}$  & $8.88\pm0.25$           & $9.35\pm0.91$ & 28 Sep 2000 \\
\vspace*{4pt}
IRAC-1            & $3.550\pm0.375$            & $4.15\pm0.19$           & \nodata &  18 Aug 2005 \\
\vspace*{4pt}
IRAC-2            & $4.493\pm0.508$            & $2.92\pm0.14$           & \nodata &  18 Aug 2005 \\
\vspace*{4pt}
IRAC-3            & $5.731\pm0.713$            & $1.98\pm0.12$           & \nodata &  18 Aug 2005 \\
\vspace*{4pt}
IRAC-4            & $7.872\pm1.453$            & $1.22\pm0.10$           & \nodata &  18 Aug 2005 \\
\vspace*{4pt}
IRS-PUI-blue      & $15.8^{+2.9}_{-2.5}$       & $0.56\pm0.08$           & \nodata &  16 Jan 2006 \\
\vspace*{4pt}
MIPS-24           & $23.675^{+2.425}_{-2.875}$ & $0.43\pm0.07$           & \nodata &  21 Feb 2006 
\vspace*{4pt}
\enddata
\tablenotetext{a}{Flux densities dereddened using the UV--optical--IR 
extinction law from \citet{fitzpatrick07} with $E(B-V)=0.22$.  
See text for details.}
\tablenotetext{b}{Note that \citet{witherick03} mistakenly give the 
date of the {\em IUE} spectrum from which these measurements were 
taken as 1984.}
\end{deluxetable}

\begin{deluxetable}{llllllllll}
\tablewidth{0pt}
\tabletypesize{\scriptsize}
\rotate
\tablecaption{Photometry of V592 Cas \label{t:oldphot}} 
\tablehead{
\colhead{JD} & 
\colhead{Year} & 
\colhead{$U$} & 
\colhead{$B$} & 
\colhead{$V$} & 
\colhead{$R$} & 
\colhead{$J$} & 
\colhead{$H$} & 
\colhead{$K_{\rm s}$} & 
\colhead{Reference}
}
\startdata
c.\ 2438761            & c.\ 1965           & 12.07:  & 12.86:     & 12.79:                    & \nodata                  & \nodata                  & \nodata                  & \nodata                  & \cite{haug70} \\
2449394.20--2449401.28 & 1993.112--1994.131 & \nodata & 12.5--12.7 & \nodata                   & \nodata                  & \nodata                  & \nodata                  & \nodata                  & \cite{taylor98} \\
2450302.68--2450303.72 & 1996.600--1996.603 & \nodata & \nodata    & 12.8(1)                   & \nodata                  & \nodata                  & \nodata                  & \nodata                  & \cite{huber98} \\
2450357.84             & 1996.751           & \nodata & \nodata    & \nodata                   & \nodata                  & 12.8(2)\tablenotemark{a} & 12.2(2)\tablenotemark{a} & 12.3(2)\tablenotemark{a} & \cite{huber98} \\
2450741.64--2450836.62 & 1997.803--1998.063 & \nodata & \nodata    & 12.7(1)                   & \nodata                  & \nodata                  & \nodata                  & \nodata                  & \cite{taylor98} \\
2451815.92             & 2000.742           & \nodata & \nodata    & \nodata                   & \nodata                  & 12.294(23)               & 12.256(31)               & 12.189(23)               & 2MASS \\
2452568.53--2452568.67 & 2002.805           & \nodata & \nodata    & \nodata                   & 12.16--12.46             & \nodata                  & \nodata                  & \nodata                  & \cite{kato02} \\
2453651.87--2453651.95 & 2005.770           & \nodata & \nodata    & 12.55(10)\tablenotemark{b} & \nodata                  & \nodata                  & \nodata                  & \nodata                  & \cite{kafka08} \\
2453742.68--2454450.62 & 2006.019--2007.959 & \nodata & \nodata    & \nodata                   & 12.3(3)\tablenotemark{c} & \nodata                  & \nodata                  & \nodata                  & AAVSO\tablenotemark{d}
\enddata
\tablenotetext{a}{We have corrected the near-IR photometry listed in 
Table 2 of \citet{huber98} by using the 2MASS photometry of their 
comparison stars C(IR) and K(IR).  These stars have 2MASS $JHK_{\rm s}$ 
magnitudes of 14.041(48), 13.772(62), 13.566(57) for C(IR) and 
14.898(40), 14.304(58), 14.186(69) for K(IR), respectively.  
The 2MASS values are consistently offset by 1.0, 0.3, and 1.1 mag 
in $J$, $H$, and $K_{\rm s}$, respectively, compared to the 
comparison star photometry listed in \citet{huber98} (and also 
used by \citealt{ciardi98} and \citealt{taylor98}.}
\tablenotetext{b}{Mean $V$ magnitude; the variations in the light 
curve range between $V=12.4$--$12.65$.}
\tablenotetext{c}{Photometry is ``Visual'', which is approximately $R$-band.}
\tablenotetext{d}{Henden, A.~A., 2008, Observations from the AAVSO 
International Database, private communication.}
\end{deluxetable}

\begin{deluxetable}{llll}
\tablewidth{0pt}
\tabletypesize{\scriptsize}
\rotate
\tablecaption{Model component parameters \label{t:model}} 
\tablehead{
\colhead{Component} & 
\colhead{Parameter} & 
\colhead{Value} &
\colhead{Source}
}
\startdata
System: & Inclination, $i$ ($^{\circ}$) & $28^{+11}_{-10}$ & \citet{huber98} \\
        & Orbital Period, $P_{\rm orb}$ (d) & 0.115063(1) & \citet{taylor98} \\
        & Distance, $d$ (pc) & 364 & \citet{taylor98}, this work \\
WD:     & Temperature, $T_{\rm wd}$ (K) & 45,000 & this work \\
        & Mass, $M_{\rm wd}$ ($M_{\odot}$) & 0.75 & this work \\
        & Radius, $R_{\rm wd}$ ($R_{\odot}$) & $0.0106$ & this work \\
SS:     & Spectral Type & M5.0 & \citet{smith98,knigge06} \\
        & Temperature, $T_2$ (K) & 3030 & \citet{cox00} \\
        & Mass, $M_2$ ($M_{\odot}$) & 0.21 & \citet{cox00} \\
ACD:    & Inner Radius, $R_{\rm acd,in}$ ($R_{\rm wd}$) & 1.0, 1.0 & this work \\
        & Outer Radius, $R_{\rm acd,out}$ ($R_{\rm wd}$) & 35.0, 35.0 & this work \\
        & Critical Radius, $R_{\rm crit}$ ($R_{\rm wd}$) & n/a, 12.25 & this work \\
        & Inner Height\tablenotemark{a}, $h_{\rm acd}$ ($R_{\rm wd}$) & 0.1, 0.1 & this work \\
        & Outer Height\tablenotemark{a}, $h_{\rm acd}$ ($R_{\rm wd}$) & 0.1, 3.36 & this work \\
        & Inner Temperature Profile Exponent, $\gamma$ & 0.75, 0.75 & this work \\
        & Outer Temperature Profile Exponent, $\gamma_{\rm out}$ & n/a, 0.70 & this work \\
        & Mass Transfer Rate, $\dot{M}$ ($M_{\odot}$ yr$^{-1}$) & $1.1\times10^{-8}$, $1.5\times10^{-8}$ & this work \\
CBD:    & Optical Depth Prescription & thin & this work \\
        & Temperature Profile Exponent & 0.75 & this work \\
        & Constant Height, $h_{\rm cbd}$ ($R_{\rm wd}$) & 0.01 & this work \\
        & Grain Density, $\rho_{\rm grain}$ (g cm$^{-3}$) & 3.0 & this work \\
        & Grain Radius, $r_{\rm grain}$ ($\mu$m) & 1 & this work \\
        & Inner Radius, $R_{\rm cbd,in}$ ($R_{\rm wd}$) & 700 & this work \\
        & Outer Radius, $R_{\rm cbd,out}$ ($R_{\rm wd}$) & 15,000 & this work \\
        & Inner Temperature, $T_{\rm cbd,in}$ (K) & 500 & this work \\
        & Outer Temperature, $T_{\rm cbd,in}$ (K) & 50 & this work \\
        & Total Mass, $M_{\rm cbd}$ (g) & $2.3\times10^{21}$ & this work \\
\enddata
\tablenotetext{a}{The disk height is the full thickness of the disk; that is, twice the distance from the disk midplane to its face.}
\tablecomments{WD = white dwarf, SS = secondary star, ACD = accretion disk, 
CBD = circumbinary disk.  Two values are listed for each of the ACD parameters, corresponding to the two models shown in Figure \ref{f:sed}; all other parameters are the same for both models.}
\end{deluxetable}



\clearpage

\end{document}